\documentclass[12pt,preprint]{aastex}

\slugcomment{to appear in ApJ, September 20, 2003}

\shortauthors{Nishikawa et al.}

\begin{document}

\title{Particle Acceleration in Relativistic Jets due to Weibel Instability}

\author{K.-I. Nishikawa\altaffilmark{1}}
\affil{National Space Science and Technology Center,
  Huntsville, AL 35805}
\email{ken-ichi.nishikawa@nsstc.nasa.gov}
\author{P. Hardee}
\affil{Department of Physics and Astronomy,
  The University of Alabama, 
  Tuscaloosa, AL 35487}
\author{G. Richardson\altaffilmark{1}}
\affil{National Space Science and Technology Center,
  Huntsville, AL 35805}
\author{R. Preece}
\affil{Department of Physics, 
  University of Alabama in Huntsville,
  Huntsville, AL 35899 and National Space Science and Technology Center,
  Huntsville, AL 35805}
\author{H. Sol}
\affil{LUTH, Observatore de Paris-Meudon, 5 place Jules Jansen 92195 
   Meudon Cedex, France}
\and
\author{G. J. Fishman}
\affil{NASA-Marshall Space Flight Center,
National Space Science and Technology Center,
  320 Sparkman Drive, SD 50, Huntsville, AL 35805}

\altaffiltext{1}{NRC Associate / NASA Marshall Space Flight Center}

\begin{abstract}

Shock acceleration is an ubiquitous phenomenon in astrophysical plasmas. 
Plasma waves  and their associated instabilities (e.g., the Buneman 
instability, two-streaming instability, and the
Weibel instability) created in the shocks are responsible for particle 
(electron, positron, and ion) acceleration. Using a 3-D relativistic
electromagnetic particle (REMP) code, we have investigated particle
acceleration associated with a relativistic jet front propagating
through an ambient plasma with and without initial magnetic fields. We
find only small differences in the results between no ambient and weak
ambient magnetic fields. Simulations show that the Weibel instability
created in the collisionless shock front accelerates particles
perpendicular and parallel to the jet propagation direction. While some 
Fermi acceleration may occur at the jet front, the majority of electron
acceleration takes place behind the jet front and cannot be characterized
as Fermi acceleration. The simulation results show that this instability 
is responsible for generating and amplifying highly nonuniform, 
small-scale magnetic fields, which
contribute to the electron's transverse deflection behind the jet
head.  The ``jitter'' radiation (Medvedev 2000) from deflected electrons
has different properties than synchrotron radiation which is calculated
in a uniform magnetic field. This jitter radiation may be
important to understanding the complex time evolution and/or spectral
structure in gamma-ray bursts, relativistic jets, and supernova
remnants.
\end{abstract}


\section{Introduction}

Nonthermal radiation, usually having a power-law emission spectrum, 
have been observed from astrophysical systems 
containing relativistic jets, e.g., active galactic nuclei (AGNs), gamma-ray 
bursts (GRBs), Galactic microquasar systems, and Crab-like supernova remnants 
(SNRs). In most of these systems, the emission is thought to be
generated by accelerated electrons through the 
synchrotron or inverse Compton mechanisms. Radiation is observed from these
systems from the radio through the gamma-ray region.

	The most widely known mechanism for the acceleration of particles 
in astrophysical environments characterized by a power-law spectrum 
is Fermi acceleration. This mechanism for particle 
acceleration relies on the shock jump conditions at relativistic shocks 
(e.g., Gallant 2002). Most astrophysical shocks are collisionless since 
dissipation is dominated by wave-particle interactions rather than 
particle-particle collisions. Diffusive shock acceleration (DSA) relies on 
repeated scattering of charged particles by magnetic irregularities 
(Alfv\'en waves) to confine the particles near the shocks. However, particle 
acceleration near relativistic shocks is not due to DSA because the 
propagation of accelerated particles ahead of 
the shock cannot be described as spatial diffusion. Anisotropies in the
angular distribution of the accelerated particles are large, and the 
diffusion approximation for spatial transport does not apply (Achterberg 
et al. 2001). Particle-in-cell (PIC) simulations may shed light on the physical 
mechanism of particle acceleration that involves the complicated dynamics 
of particles in relativistic shocks.

Recent PIC simulations using counter-streaming relativistic jets show that
acceleration is provided in situ in the downstream jet, rather than by the 
scattering of particles back and forth across the shock as in Fermi 
acceleration (Frederiksen et al. 2003).  Two independent simulation 
studies have confirmed that the counter-streaming jets excite the Weibel 
instability (Weibel 1959), which generates magnetic fields (Medvedev \& 
Loeb 1999; Brainerd 2000; Pruet et al. 2001; Gruzinov 2001), and 
consequently accelerates electrons (Silva et al. 2003; Frederiksen et 
al. 2003).  Since these simulations are performed with counter-streaming 
jets, shock dynamics involving the jet head, where Fermi acceleration may 
take place, has not been investigated.

   	In this paper we present new simulation results of particle 
acceleration and magnetic field generation in relativistic jets using 3-D
relativistic electromagnetic particle-in-cell (REMP) simulations with and 
without  initial ambient magnetic fields. In our simulations, an electron-ion 
relativistic jet with Lorentz factor, $\gamma = 5$ (corresponds to 5 MeV) 
is injected into an electron-ion plasma in order to study the dynamics of 
a relativistic collisionless 
shock. We illustrate the features of the collisionless shock generated at 
the head of the relativistic jet injected into magnetized and unmagnetized 
ambient plasma. The Weibel instability is excited in the downstream region 
and accelerates electrons and ions. In section 2 the simulation model and 
initial conditions are described. The simulation results are presented in 
the section 3, and in section 4 we summarize and discuss the results.

\section{Simulation model}

  	The code used in this study is a modified version of the TRISTAN code, 
which is a relativistic electromagnetic particle (REMP) code (Buneman 1993). 
Descriptions of PIC codes are presented in Dawson 
(1983), Birdsall \& Langdon (1995), and Hickory \& Eastwood (1988). This code 
has been used for many applications including astrophysical plasmas (Zhao 
et al. 1994; Nishikawa et al. 1997).

The simulations were performed using $85 \times 85 \times 160$ grids with 
a total of 55 to 85 million particles (27 particles$/$cell$/$species for 
the ambient plasma). Both periodic and radiating boundary conditions are 
used (Buneman 1993). The ambient electron and ion plasma has mass ratio 
$m_{\rm i}/m_{\rm e} = 20$. The electron thermal velocity $v_{\rm e}$ is 
$0.1c$, where $c$ is the speed of light. The electron skin depth, 
$\lambda_{\rm ce} = c/\omega_{\rm pe}$, is $4.8\Delta$, where 
$\omega_{\rm pe} = (4\pi e^{2}n_{\rm e}/m_{\rm e})^{1/2}$ is the electron 
plasma frequency ($\Delta$ is the grid size). In this study three different 
cases are simulated to investigate the fundamental characteristics of the 
Weibel instability in electron-ion plasmas. One simulation is performed 
injecting a thin jet into a magnetized ambient plasma. The other two 
simulations consider flat jets (infinite width) injected into a magnetized 
ambient plasma and then into an unmagnetized ambient plasma in order to 
compare this simulation with previous simulations (Silva et al. 2003; 
Frederiksen et al. 2003).

\section{Simulation results}
\subsection{Injection into Magnetized Ambient Plasma}
\subsubsection{Thin jet}

      	Two kinds of jets have been simulated: a ``thin'' jet with 
radius $r_{\rm jet} = 4\Delta$ and a ``flat" (thick) jet that fills the 
computational domain in the transverse directions (infinite width). In 
the simulations, jets are injected at $z = 25\Delta$. First, we describe 
the simulation results for the thin jet with initial ambient magnetic 
fields.
The number density of the thin jet is $0.741n_{\rm b}$, where $n_{\rm b}$ is 
the density of ambient (background) electrons.  The average jet velocity 
$v_{\rm j} = 0.9798c$, and the Lorentz factor is 5. The time step 
$\omega_{\rm pe}t = 0.026$, the ratio 
$\omega_{\rm pe}/\Omega_{\rm e} = 2.89$, the Alfv\'en speed 
$v_{\rm A} = 0.0775c$, and the Alfv\'en Mach number  
$M_{\rm A} = v_{\rm j}/v_{\rm A} = 12.65$. The gyroradii of ambient 
electrons and ions are $1.389\Delta$, and $6.211\Delta$, respectively. As shown in 
Fig. 1 the jet electrons are twisted due to a two-stream-type instability.
Since the size of jet is of the order of the electron skin depth, it is 
not clear if the Weibel instability is excited. As will be shown in the 
flat jet simulation, the density perturbation is similar to that induced by the 
Weibel instability. Since the size of the jet is extremely small 
compared to realistic jets, further study of thin jets will be performed 
including simulations with larger jet sizes.

\subsubsection{Flat jet} 

    	The jet density of the flat jet is also nearly $0.741n_{\rm b}$. 
Other parameters are the same as in the case of the thin jet. In this case,
the jet makes contact with the ambient plasma at a 2D interface spanning 
the computational domain. Therefore, only the dynamics of the jet head and 
the propagation of a shock in the downstream region is studied. This 
simulation system is similar to simulations performed previously using 
effectively infinite counter-streaming jets. The important difference between 
this simulation and previous simulations is that the evolution of the Weibel 
instability is examined in a more realistic spatial way including motion 
of the jet head. Furthermore, the density of the jet relative to the 
ambient plasma can be changed easily.
     	The Weibel instability is excited and the electron density is 
perturbed as shown in Fig. 2a. The electrons are deflected by the perturbed 
(small) transverse magnetic fields ($B_{\rm x}, B_{\rm y}$) via the Lorentz 
force: $-e({\bf v} \times {\bf B})$, generating filamented current 
perturbations ($J_{\rm z}$), which enhance the transverse magnetic fields 
(Weibel 1959; Medvedev and Loeb 1999). The complicated current structures 
due to the Weibel instability are shown in Fig. 2b. The sizes of these 
structures are nearly the electron skin depth ($4.8\Delta$). This is in good 
agreement with 
$\lambda \approx 2^{1/4}c\gamma_{\rm th}^{1/2}/\omega_{\rm pe} \approx 
1.188\lambda_{\rm ce} = 5.7\Delta$ (Medvedev \& Loeb 1999). Here, 
$\gamma_{\rm th}$ is a thermal Lorentz factor, $\omega_{\rm pe}$ is the 
electron plasma 
frequency. The shapes are elongated along the direction of the jet 
(the $z$-direction, horizontal in Fig. 2). 
The acceleration of electrons has been reported in previous work 
(Silva et al. 2003; Frederiksen et al. 2003). As shown in Fig. 3, the 
kinetic energy (parallel velocity $v_{\parallel} \approx v_{\rm j}$) of 
the jet electrons is transferred to the perpendicular velocity via the 
electric and magnetic fields generated by the Weibel instability. The 
strongest deceleration of electron flow (Fig. 3b) and strongest 
transverse acceleration (Fig. 3a) is between $z/\Delta = 100 - 120$ and 
takes place around the maximum amplitude of perturbations due to the 
Weibel instability at $z = 112\Delta$ (see Figs. 2 and 4). On the other 
hand, jet ions are slightly accelerated (not shown). However, due to the 
larger mass of the ions, the energy gained by ion acceleration is similar 
to that gained by the electrons. Furthermore, at the jet front some jet 
electrons are accelerated and some are decelerated. This acceleration and 
deceleration is shown by the slanting in the parallel velocity distribution 
at the jet head (Fig. 3b). This may indicate that Fermi acceleration is 
taking place at the jet front as described in previous work (e.g., 
Achterberg et al. 2001; Gallant 2002; Ellison \& Double 2002).

      The Weibel instability is excited just behind the jet head as shown 
in Fig. 4. The three curves are measured at three different locations 
($y/\Delta = 38, 43$, and 48) about the electron skin depth apart. 
As a result the phase of instability is different, but the amplitudes 
are similar. The growth rate of the Weibel instability is calculated to 
be, $\tau \approx \gamma_{\rm sh}^{1/2}/\omega_{\rm pe} \approx 21.4$ 
($\gamma_{\rm sh} = 5$) (Medvedev \& Loeb 1999). This 
is in good agreement with the simulation results with the jet head located 
at $z = 136\Delta$. Figure 3 suggests that the ``shock" has a thickness from 
about $z/\Delta = 80 - 130$ behind the shock front. Possibly, the 
``turbulence" assumed for the diffusive shock acceleration (DSA) corresponds 
to this shock region. The transverse acceleration of jet electrons takes 
place in the shock, as shown in Fig. 3a. Subtle structures are recognized 
at the jet head ($z = 136\Delta$). As shown in Fig. 4a, the width of the jet 
head is nearly the electron skin depth ($4.8\Delta$). Additionally, Figure 4b shows
that the $z$-component of current density becomes slightly negative. This 
results from the fact that some jet electrons are ahead of the ions. This 
negative current at $z = 137\Delta$ is clearly shown in Fig. 5b. Based on 
Figs. 2 and 4 the size of perturbations along the jet around $z = 120\Delta$ 
is nearly twice the electron skin depth. This result is consistent with 
the previous simulations by Silva et al. (2003).

      The Weibel instability is excited just behind the jet head located at 
$z = 136\Delta$ (see Fig. 5c ($z = 134\Delta$)), and the accelerated jet 
electrons are creating the negative current at $z = 137\Delta$ (Fig. 5b) 
just in front of the jet head. This total negative current is indicated by 
three curves dipped uniformly at 
$z = 137\Delta$ in Fig. 4b. 

     	The Weibel instability creates elongated shell-type structures which 
are also shown in counter-streaming jet simulations (Silva et al. 2003; 
Frederiksen et al. 2003). The size of these structures transverse to the 
jet propagation is nearly the electron skin depth ($4.8\Delta$) as shown in 
Figs. 5 and 6.  It should be noted that the size of perturbations at 
$z = 120\Delta$ is larger than those at different locations, 
since the smaller scale perturbations merge to larger sizes in the 
nonlinear stage at the maximum amplitudes (Silva et al. 2003).

\subsection{Injection into Unmagnetized Ambient Plasma (Flat Jet)}

    In the previous simulations of Silva et al. (2003) and Frederiksen et 
al. (2003) ambient magnetic fields are not included initially. In order to 
investigate the generation of magnetic fields, a simulation has been 
performed without ambient magnetic fields but otherwise with the same 
parameters.

	The structure of perturbations to the electron density and 
$z$-component of the current density are similar to those with initial
 ambient magnetic fields ($B_{0z}$), as shown in Fig. 7.  
The unmagnetized jet generates magnetic fields due to the  
current structures produced in the Weibel instability, as shown in Figs. 
7b and 8b. The 
peak values of the perturbations due to the Weibel instability for an 
unmagnetized ambient plasma is larger that those for a magnetized ambient 
plasma. However, the amplitude of these perturbations is similar; therefore 
the effects of initial ambient magnetic fields are not apparent in these 
cases. Further investigation is required for a systematic analysis of the 
effects of magnetic fields. 
     The generation of magnetic fields both with and 
without an initial magnetic field suggests that synchrotron emission (or
jitter radiation) is relevant in GRB afterglows and Crab-like pulsar winds 
(Medvedev 2000).

	Figure 9 shows the structures of perturbations along the $z$-direction 
in electron density, $J_{\rm z}$, $E_{\rm z}$, and $B_{\rm x}$, without an 
initial magnetic field. These
are very similar to those with an initial magnetic field, shown in Fig. 4. 
The three curves are measured at three different locations ($y/\Delta = 38, 
43$, and 48) each separated about one electron skin depth. The phase of the 
instability is different along each line, but the amplitudes are similar. 
Furthermore, the different seed perturbations (thermal noise by the initial
loading of particles) and the unmagnetized plasma make the amplitudes and shape 
of perturbations at the nonlinear phase different from those with the 
magnetized plasma (as shown in Fig. 4). From the
previous simulations without an initial ambient magnetic field (Silva et al.
2003; Frederiksen et al. 2003), highly nonuniform, small-scale magnetic 
fields are generated due to the Weibel instability.

\section{Summary and Discussions}

     	 We have performed the first self-consistent, three-dimensional 
relativistic particle simulations of  electron-ion relativistic jets 
propagating through magnetized and unmagnetized electron-ion ambient 
plasmas. The Weibel instability is excited in the downstream region behind 
the jet head, where electron density perturbations and filamented currents 
are generated. The nonuniform electric field and magnetic field structures 
slightly decelerate the jet electrons and ions, while accelerating (heating) 
the jet electrons and ions in the transverse direction, in addition to 
accelerating the ambient material. The origin of the Weibel instability 
relies on the 
fact that the electrons are deflected by the perturbed (small) transverse 
magnetic fields ($B_{\rm x}, B_{\rm y}$), and subsequently enhance the 
filamented current  (Weibel 1959; Medvedev and Loeb 1999; Brainerd 2000; 
Gruzinov 2001). The deflection of particle orbits due to the Lorentz force 
increases as the magnetic field perturbation grows in amplitude. The 
amplified magnetic field is random in the direction perpendicular to the 
particle motion, since it is generated from a random seed field. The 
perturbed electron density and filamented currents have a complicated 
three-dimensional structure. The transverse size of these structures is 
nearly equal to the electron skin depth but is larger if there are no ambient 
magnetic fields. However, the size along the direction of jets is larger 
than the transverse scales. At the termination of our simulation, the 
thickness of the unstable region along the jet direction ranges from 
$z/\Delta = 80$ to $130$. The perturbation size in the transverse direction 
become largest around $z/\Delta = 120$, where nonlinear effects lead to 
the merging of the smaller scale filaments that first appear behind the jet 
front. This result is similar to previous counter-streaming simulation 
results (Silva et al. 2003) in which smaller scale filaments first appear 
and then merge into larger scale filaments at a later time. Now 
we see the temporal development appear as a spatial development.

	The simulation results show that the initial jet kinetic energy goes 
to the magnetic fields and transverse acceleration of the jet particles through 
the Weibel instability. The properties of the synchrotron or ``jitter" 
emission from relativistic shocks are determined by the magnetic field 
strength, ${\bf B}$ and the electron energy distribution behind the shock. 
The following dimensionless parameters are used to estimate these values; 
$\epsilon_{\rm B} = U_{\rm B}/e_{\rm th}$ and $\epsilon_{\rm e} = 
U_{\rm e}/e_{\rm th}$ (Medvedev \& Loeb 1999). Here $U_{\rm B} = B^2/8\pi$, 
$U_{\rm e}$ are the magnetic and electron energy densities, and 
$e_{\rm th} = nm_{\rm i}c^{2}(\gamma_{\rm th} - 1)$ is 
the total thermal energy density behind the shock, where $m_{\rm i}$ is 
the ion mass, $n$ is the ion number density, and $\gamma_{\rm th}$ is the mean 
thermal Lorenz factor of ions. Based on the available diagnostics the following 
values are estimated; $\epsilon_{\rm B} \approx 0.02$ and $\epsilon_{\rm e}
\approx 0.3$ (typical values for GRB afterglows). These estimates are made at 
the maximum amplitude ($z \approx 112\Delta$).

	The basic nature of the Weibel instability is as follows 
(Medvedev \& Loeb 1999):

\begin{list}{}{\setlength{\leftmargin}{0.5cm} \setlength{\itemindent}{-0.5cm}}
\item
\vspace{-1.0cm}
\item
1.	The instability is aperiodic, i.e., $\omega_{\rm real}$ = 0 (convective). 
Thus, it can be saturated only by nonlinear effects and not by kinetic effects, 
such as collisionless damping or resonance broadening. Hence the magnetic 
field can be amplified to very high values. 
\item
2.	The instability is self-saturating. It continues until all the 
free energy due to the particle distribution function (PDF) anisotropy 
is transferred to the magnetic field energy. 
\item
3.	The produced magnetic field is randomly oriented in the direction
perpendicular to 
the flow. The Lorentz forces randomize particle motion over the pitch 
angle, and hence introduce an effective scattering into the otherwise 
collisionless system.
\end{list}

To this basic nature we add the spatial development associated with 
the propagation of the jet front:

\begin{list}{}{\setlength{\leftmargin}{0.5cm} \setlength{\itemindent}{-0.5cm}}
\item
\vspace{-1.0cm}
\item
1'.	The smaller scale filaments appear immediately behind the jet front. 
The nonlinear effect is recognized by the fact that smaller filaments 
merge into larger filaments behind the jet front, as shown in Figs. 5 and 
6.
\item
2'.	The instability is self-saturating (spatially) behind the jet front 
and continues until all of the free energy due to the particle distribution 
function anisotropy is transferred to the magnetic field energy. 
The largest amplitude of perturbations is located near the region with 
the maximum transverse acceleration (Fig. 3a) with deceleration in 
parallel velocity ($v_{\rm z}$) (Fig. 3b), as shown in Figs. 2, 5d, and 6d. 
\item
3'. The produced magnetic field is randomly oriented in the ``shock" 
plane (three-dimensional structure, as shown in Figs. 2b and 5). 
The random Lorentz forces result in randomization and deceleration of the 
jet particles, an acceleration of the ambient particles, and transverse 
acceleration (heating) of jet and ambient particles inside the ``shock'' 
region.
\end{list}

        Previous microphysical analyses of the energy conversion in
relativistic pair outflows interacting with the surrounding interstellar 
medium consisting of cold protons and electrons have been performed 
(e.g., Brainerd 2000; Schlickeiser et al. 2002). These analyses have 
demonstrated that the beam excites both electrostatic and low-frequency 
magnetohydrodynamic Alfv\'en-type waves via a two-stream instability in the 
background plasma. They have also provided the time evolution of the distribution 
functions of beam particles and the generated plasma wave turbulence power 
spectra. While the jet front showed evidence of Fermi acceleration, the 
main acceleration of electrons may take place in the downstream region 
(e.g., Brainerd 2000; Schlickeiser et al. 2002; Ostrowski \& Bednarz 2002). 

	Our present simulation study has provided the framework of the
fundamental dynamics of a relativistic shock generated within a relativistic 
jet. While some Fermi 
acceleration may occur at the jet front, the majority of electron 
acceleration takes place behind the jet front and cannot be characterized 
as Fermi acceleration.  Since the shock dynamics is complex and subtle, 
further comprehensive study is required for better understanding of the 
acceleration of electrons and the associated emission as compared with 
current theory (e.g., Rossi \& Rees 2002).  This further study will provide more 
insight into basic relativistic collisionless shock characteristics.  
The fundamental characteristics of such shocks are essential for a proper 
understanding of the prompt gamma-ray and afterglow emission in gamma-ray 
bursts, and also to an understanding of the particle reacceleration 
processes and emission from the shocked regions in relativistic AGN 
jets.

\acknowledgments
K.N. is a NRC Senior Research Fellow at NASA Marshall Space Flight Center.
This research (K.N.) is partially supported by NSF ATM 9730230,
ATM-9870072, ATM-0100997, and INT-9981508. The simulations 
have been performed on ORIGIN 2000 and IBM p690 (Copper) at NCSA which 
is supported by NSF.

\clearpage

\begin{figure}[ht]
\epsscale{1.00}
\vspace*{-2.0cm}
\plotone{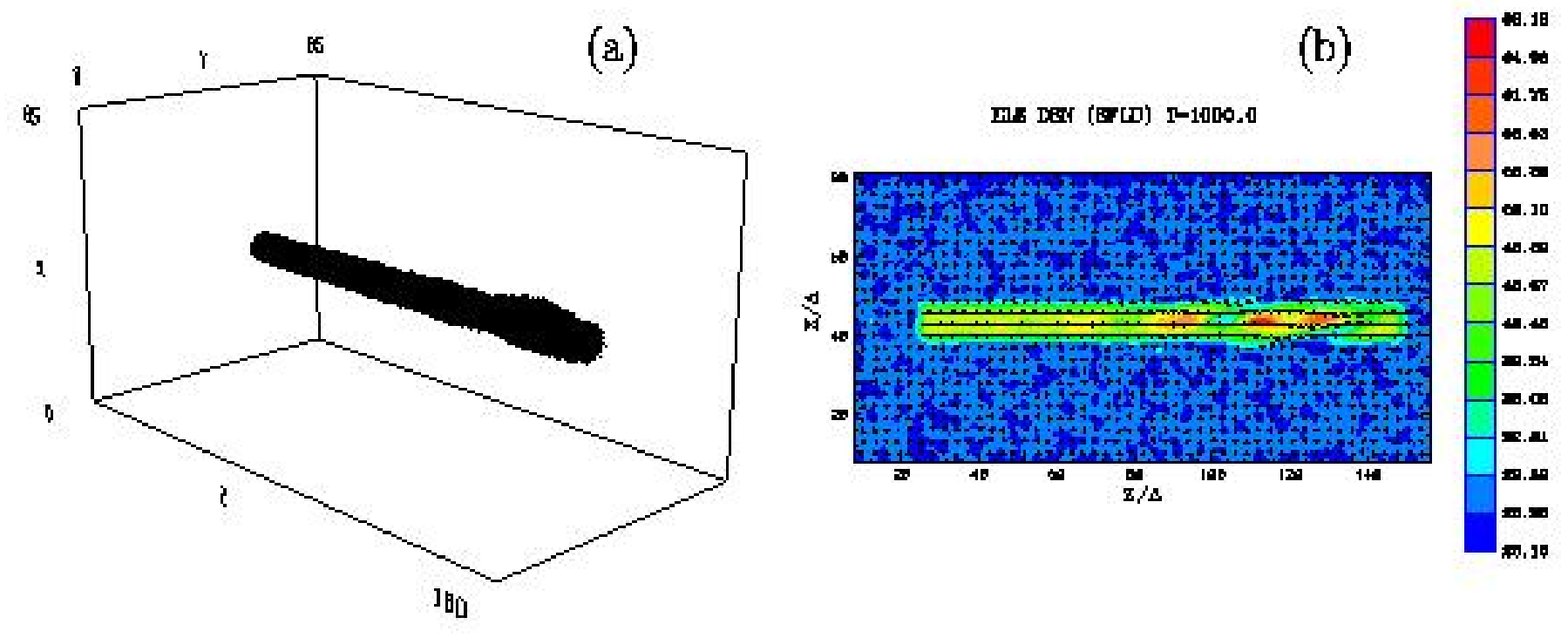}
\vspace*{-11.0cm}
\caption{Dynamics of the thin jet are indicated at $\omega_{\rm pe}t = 26$
by (a) a jet electron image in the 3-dimensional simulation system, and  
(b) the total electron density in the $x - z$ plane in the center of the jet 
with the electron flux indicated by arrows and density indicated by color. 
A two-stream-type instability twists the jet electrons as shown in these 
plates.}
\end{figure}

\begin{figure}[ht]
\epsscale{1.00}
\vspace*{-5.0cm}
\plotone{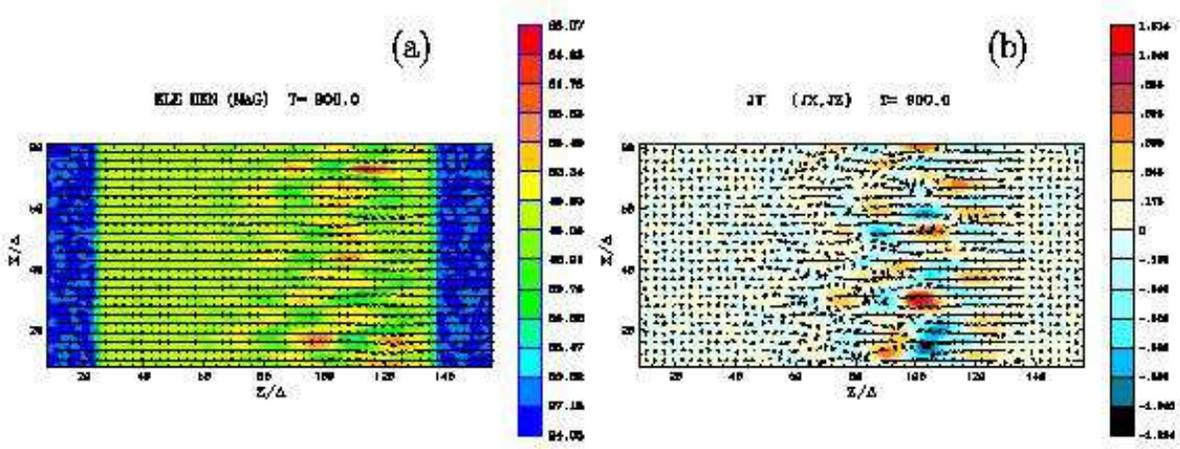}
\vspace*{-11.5cm}
\caption{The Weibel instability for the flat jet is illustrated in 2D images 
in the $x - z$ plane 
($y = 43\Delta$) in the center of the jet at $\omega_{\rm pe}t = 23.4$.  In 
(a) the colors indicate the electron density with magnetic fields represented 
by arrows and in (b) the colors indicate the $y$-component of the current 
density ($J_{\rm y}$) with $J_{\rm z}, J_{\rm x}$ indicated by the arrows. 
The Weibel instability perturbs the electron density, leading to nonuniform 
currents and highly structured magnetic fields.}
\end{figure}

\begin{figure}[ht]
\epsscale{0.95}
\vspace*{-5.0cm}
\plotone{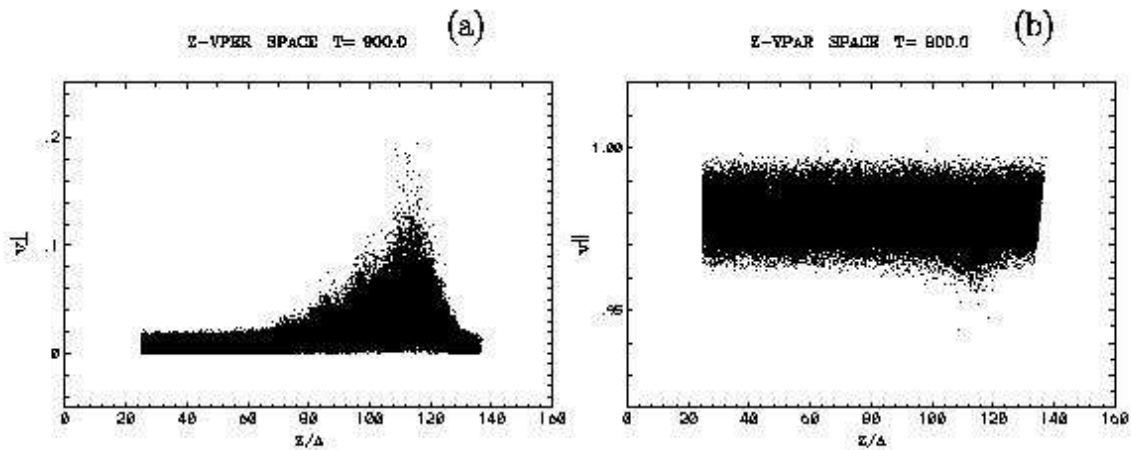}
\vspace*{-10.5cm}
\caption{The distribution of jet electrons at $\omega_{\rm pe}t = 23.4$
in (a) $z - v_{\perp}/c$ and (b) $z - v_{\parallel}/c$ phase space. Roughly 
20\% of the jet electrons are randomly selected for these plots.  
(a) Jet electrons are accelerated perpendicularly  ($v_{\perp} = (v_{\rm x}^2 
+ v_{\rm y}^2)^{1/2}$) due to the nonuniform electric and magnetic fields. 
(b) Jet electrons are decelerated while they are accelerated perpendicularly 
around $z = 112\Delta$. The parallel velocity slanting at the jet head 
($z = 136\Delta$) is due to the acceleration and deceleration, and may be 
the result of Fermi acceleration.}
\end{figure}

\begin{figure}[ht]
\epsscale{1.00}
\vspace*{-3.0cm}
\plotone{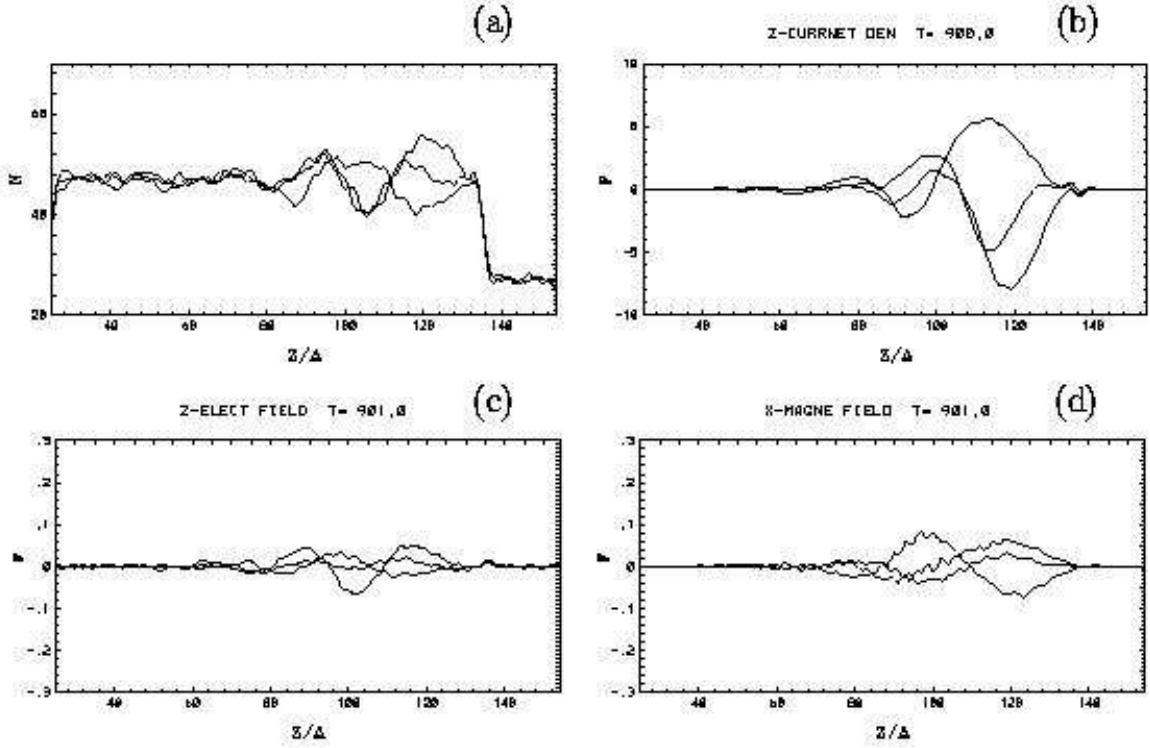}
\vspace*{-10.0cm}
\caption{One-dimensional displays along the $z$-direction 
($25 \leq z/\Delta \leq 154$) of (a) the electron density, 
(b) the $z$-component of the current density, 
(c) the $z$-component of the electric field, and (d) the $x$-component of the 
magnetic field at $\omega_{\rm pe}t = 23.4$. The three curves are obtained at 
$x/\Delta = 38$ and $y/\Delta = 38, 43$, and 48. Due to the separation by the 
electron skin depth ($\lambda_{\rm ce} \approx 4.8\Delta$) each curve shows the 
value at the different node of the growing Weibel instability.}
\end{figure}

\begin{figure}[ht]
\epsscale{0.95}
\vspace*{-2.5cm}
\plotone{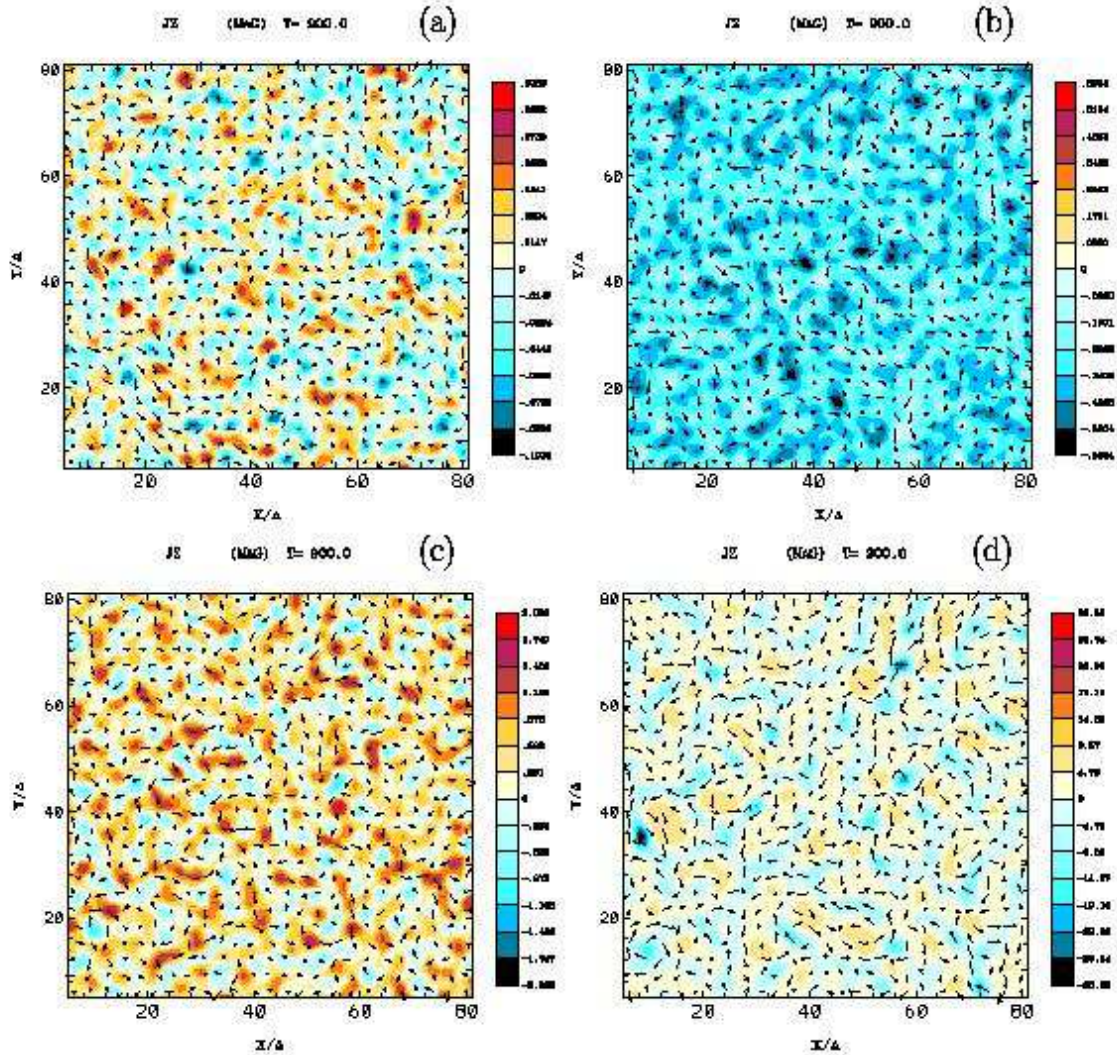}
\vspace*{-4.0cm}
\caption{The z-component of the current density in the $x - y$ plane is plotted 
at $z/\Delta =$ (a) 140, (b) 137, (c) 134, and (d) 120. The maximum amplitudes 
of these panels are (a) 0.03 (thermal noise level), (b) 0.59, (c) 2.03, and 
(d) 7.0 (peak: 33.52), respectively. The sizes of current structures are larger 
at (d) $z = 120\Delta$, where the Weibel instability grows to the maximum 
amplitude (see Fig. 4), and smaller cells have nonlinearly merged into larger 
cells.}
\end{figure}

\begin{figure}[ht]
\epsscale{0.95}
\vspace*{-2.5cm}
\plotone{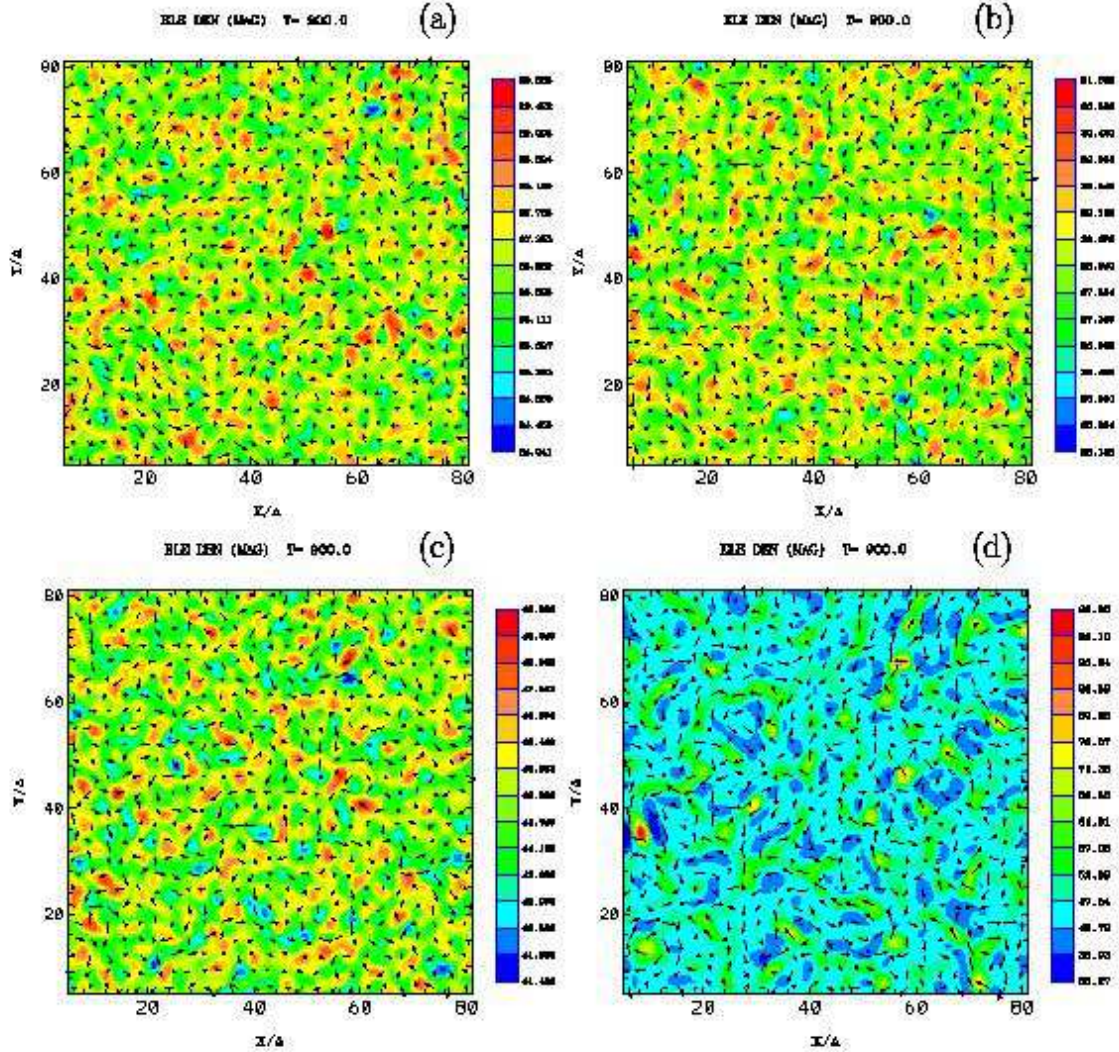}
\vspace*{-4.0cm}
\caption{The electron density in the $x - y$ plane is plotted at $z/\Delta=$ 
(a) 140, (b) 137, (c) 134, and (d) 120 (as in Fig. 5). The maximum amplitude 
in these panels is (a) 29.8  (the ambient plasma with thermal noise), (b) 31.2, 
(c) 49.2, and (d) 70 (peak: 99.8), respectively. The unperturbed average 
electron densities in the ambient and the total (jet plus ambient) plasma are 27 
and 47, respectively. The size of the density perturbation structures is larger 
around $z/\Delta = 120$ (d), since the Weibel instability grows to the maximum 
amplitude (as shown in Fig. 4a) since smaller cells are merged into larger cells 
nonlinearly. The maximum density at $z/\Delta = 137$ (b) is 31.2, thus the 
jet head is located here, as indicated by Figs. 3b and 4a.}
\end{figure}

\begin{figure}[ht]
\epsscale{1.00}
\vspace*{-5.0cm}
\plotone{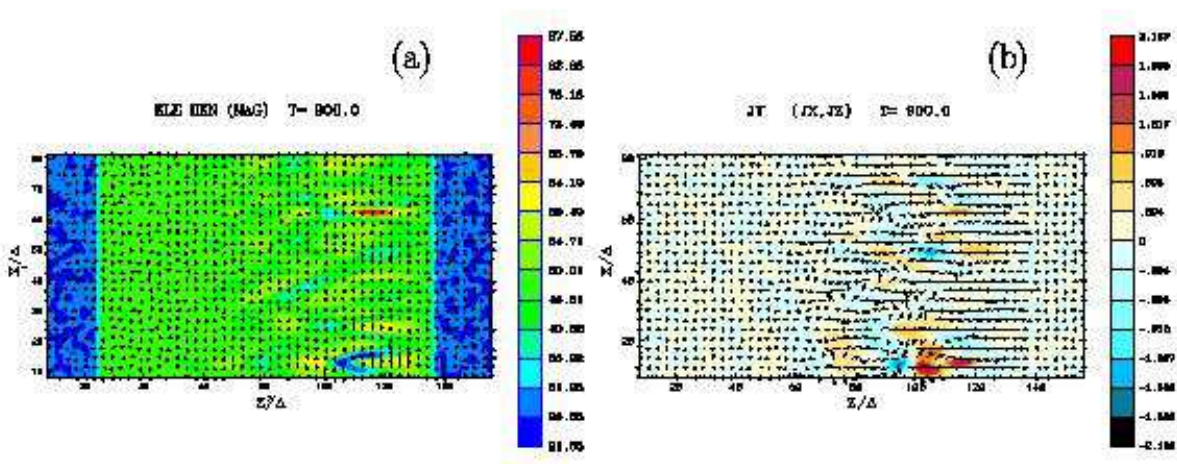}
\vspace*{-11.5cm}
\caption{For the simulation with an unmagnetized ambient plasma, the Weibel 
instability is illustrated in 2D images in the $x - z$ plane in the center of 
the jet ($y = 43\Delta$ at $\omega_{\rm pe}t = 23.4$).  In (a) the colors 
indicate the electron density with magnetic fields represented by arrows and in 
(b) the colors indicate the $y$-component of the current density $J_{\rm y}$, 
with $J_{\rm z}, J_{\rm x}$ indicated by the arrows. The Weibel instability 
perturbs the electron density, leading to nonuniform currents and highly 
structured magnetic fields. These plates correspond to those in Fig. 2, but the 
color scales are different. The patterns are similar, however, the peak values with 
an unmagnetized ambient plasma are larger than those in the magnetized case.}
\end{figure}

\begin{figure}[ht]
\epsscale{0.95}
\vspace*{-5.0cm}
\plotone{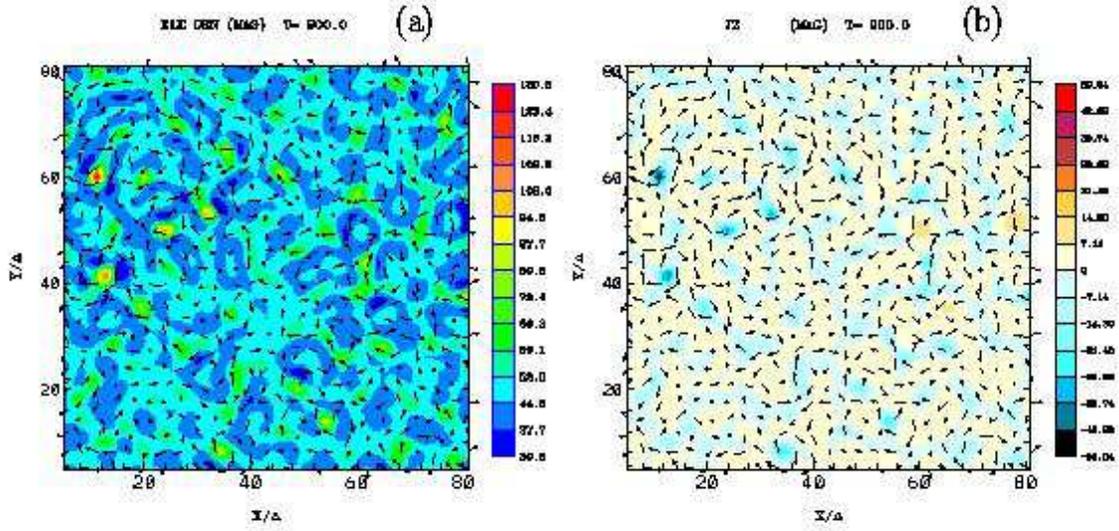}
\vspace*{-9.0cm}
\caption{The electron density (a) and $z$-component of the current density (b)
are plotted in the $x - y$ plane at $z = 120\Delta$. The maximum amplitudes of 
these panels are (a) 70 (peak: 130; compare to Fig. 6d) and (b) 10 (peak: 50;
compare to Fig. 5d). The size of the current structures is similar to those found 
with an ambient magnetic field. These plates correspond to Figs. 6d and 5d, 
respectively.}
\end{figure}

\begin{figure}[ht]
\epsscale{1.00}
\vspace*{-3.0cm}
\plotone{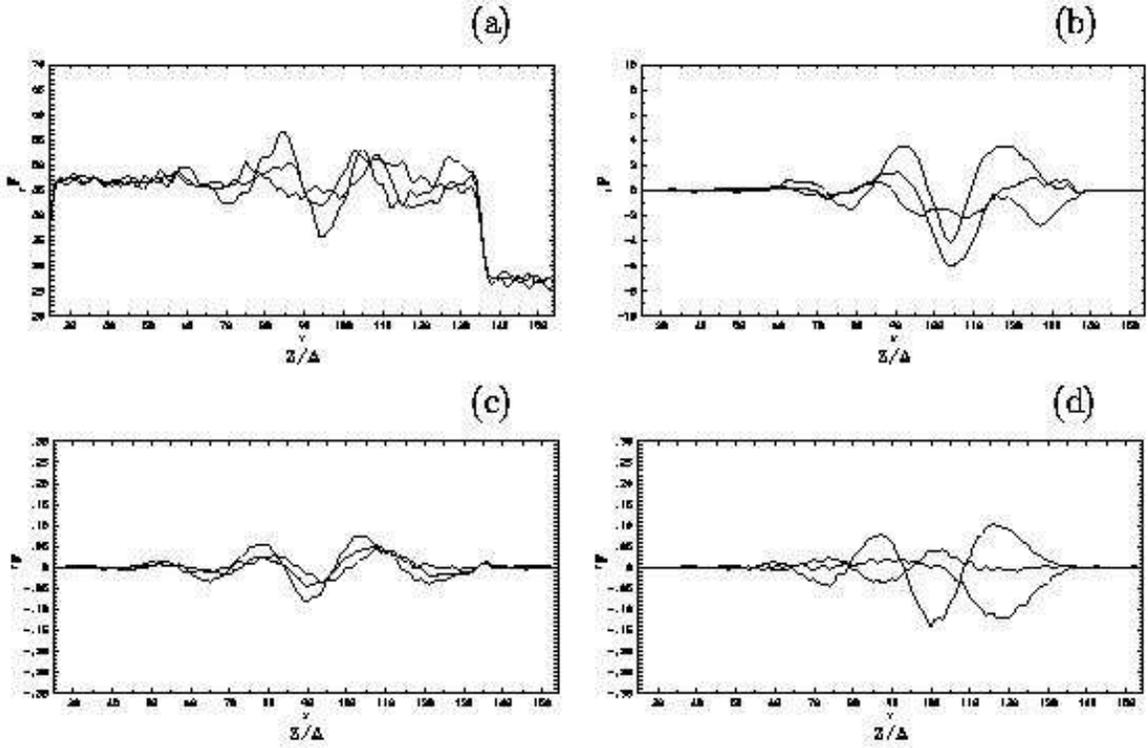}
\vspace*{-10.0cm}
\caption{One-dimensional displays along the z-direction of the electron 
density (a), the $z$-component of the current density (b), the $z$-component 
of the electric field (c), and the $x$-component of the magnetic field (d) at 
$\omega_{\rm pe}t = 23.4$. The three curves are obtained at $x/\Delta = 38$ 
and $y/\Delta = 38, 43$, and 48. These plates correspond to those in Fig. 4.}
\end{figure}

\end{document}